\documentclass{article}

\usepackage{arxiv}

\usepackage[utf8]{inputenc} 
\usepackage[T1]{fontenc}    
\usepackage{caption}
\usepackage{subcaption}
\usepackage{graphicx}
\usepackage{color}
\usepackage[section]{placeins}
\usepackage[normalem]{ulem}
\usepackage{amsmath,amsbsy,amssymb}
\usepackage{hyperref}
\usepackage{tikz}
\usepackage[title]{appendix}

\title{Assessing standard and kinetic energy conserving discontinuous Galerkin formulations for marginally resolved Navier-Stokes flows}

\author{
Bjoern F. Klose\thanks{bklose@sdsu.edu}\\
Department of Aerospace Engineering\\
San Diego State University\\
San Diego, CA 92182\\
\AND
Gustaaf B. Jacobs\thanks{Corresponding author, gjacobs@sdsu.edu}\\
Department of Aerospace Engineering\\
San Diego State University\\
San Diego, CA 92182\\
\AND
David A. Kopriva\thanks{Professor Emeritus at The Florida State University, Tallahassee, FL 32306, kopriva@math.fsu.edu}\\
Computational Science Research Center\\
San Diego State University\\
San Diego, CA 92182\\
}

\begin{document}
\maketitle

\begin{abstract}
%
%
The robustness and accuracy of marginally resolved discontinuous Galerkin spectral element computations are evaluated for the standard formulation and a kinetic energy conserving split form on complex flow problems of physical and engineering interest, including
the flow over a square cylinder, an airfoil and a plane jet.
It is shown that
the kinetic energy conserving formulation is significantly more robust than the standard scheme for under-resolved simulations.
A disadvantage of the split form is the restriction to Gauss-Lobatto nodes with the inherent underintegration and lower accuracy as compared to Gauss quadrature used with the standard scheme.
While the results support the higher accuracy of the standard Gauss form, lower numerical robustness and spurious oscillations are evident in some cases, giving the advantage to the kinetic energy conserving scheme for marginally resolved numerical simulations.
\end{abstract}


\section{Introduction} \label{introduction}
Discontinuous Galerkin spectral element methods (DGSEM) are powerful tools for conducting high fidelity numerical simulations such as direct numerical simulations (DNS) of the Navier-Stokes equations. The high-order accuracy, low memory requirement and multi-processor scaling properties make this class of methods well suited for the numerical investigation of complex flow phenomena \cite{kopriva, boyd}. 

Numerical stability is an important property for all CFD schemes. But especially for high-order spectral element methods with low inherent dissipation, the control of numerical errors becomes crucial. The analysis of fluid dynamics, such as flow instabilities and transition phenomena demands the direct computation of the Navier-Stokes equations, often in the turbulent flow regime. With higher Reynolds numbers, the computational cost to completely resolve all scales of motion becomes increasingly high, and for many applications is not feasible. 
A popular approach to overcome this issue is to numerically filter the Navier-Stokes equations with so-called large eddy simulations (LES) \cite{SJM06a}. Two LES methods are commonly used: the Implict LES (ILES) \cite{Grinstein07,KSJM10,uranga11,beck14,wiart15,flad16,moura17}, or Explictly Filtered LES (EFLES) \cite{Diamessis08,SJM09,WWVJ01}. In ILES the dissipative effect of small scale turbulence is accounted for by numerical dissipation. In EFLES an explicit numerical filter is used to numerically account for the subgrid dissipation. The use of general DGSEM for DNS can be found in \cite{AMJ14,Jacobs03,JKM03,JKM04,JKM05a}, LES in \cite{SJM06a,SJM09,SJM06} and Explicitly Filtered LES (EFLES) of turbulent flow in complex geometries using exponential filters in  \cite{KSJM10,SJM09,SSJM09,CJDAM16}.

The no-model ILES scheme is a popular choice in combination with Discontinuous Galerkin methods, as the DGSEM adds numerical viscosity through the element-interface fluxes via a dissipative Riemann solver. As a consequence, the dissipation behavior depends only on the choice of the Riemann solver, and no other user-defined sub-grid scale model parameters are needed \cite{SJM09,flad17}.

The performance of high-order methods on under-resolved turbulence simulations has been discussed by Gassner \& Beck \cite{gassner12}, with the conclusion that higher polynomial orders have superior turbulence spectrum representations compared to lower order computations with the same number of degrees of freedom (DOF). However, given the non-linear nature of the Navier-Stokes equations, energy input through aliasing errors and the lack of numerical dissipation at higher orders requires stabilization of the scheme through dealiasing, e.g. by overintegration \cite{gassner12,kirby03}.

A more recent method for stabilization of discontinuous Galerkin schemes is a split form approximation, e.g. as discussed by Gassner et al. \cite{gassner16}. The method satisfies certain conservation properties for the compressible Euler equations by manipulation of the non-linear volume fluxes using a split formulation, so that conservation of kinetic energy or entropy can be achieved. By choosing a kinetic energy conserving form of the fluxes, the scheme proves to be numerically robust. The two techniques for the non-linear advective terms, dealiasing through overintegration and split form DG schemes, are compared on simulations of under-resolved turbulence by Winters et al. \cite{winters18}, where the split form approximation has shown to have superior robustness. 

For the propagation of a linear wave, Gassner \& Kopriva \cite{gassner2011} evaluated the dissipation and dispersion errors of discontinuous Galerkin schemes using Gauss and Gauss-Lobatto quadrature nodes. They show that the Gauss variant has a higher accuracy per wavelength and beneficial dissipation properties that lead to an increased robustness in under-resolved computations. There is, however, no proof of numerical stability for the Gauss DGSEM and conservation of kinetic energy or entropy requires the split-form formulation with Gauss-Lobatto nodes \cite{gassner16}.
Therefore, one has to make a decision between the standard DGSEM scheme with Gauss nodes for higher accuracy per point or a provably stable formulation in split-form with Gauss-Lobatto nodes.


In this paper, we asses the stability and accuracy of discontinuous Galerkin approximations in the standard conservation form and in a kinetic energy conserving split form for under-resolved computations.
To this end, direct numerical simulations of a three dimensional inviscid Taylor-Green vortex, the viscous flows over a square cylinder, a plane jet, and an airfoil flow in two and three dimensions are conducted.
Following Gassner et al. \cite{gassner16}, we demonstrate the kinetic energy conservation of the DG scheme on the inviscid Taylor-Green vortex and challenge the method's robustness and accuracy by computation of the turbulent break up of a plane jet (e.g. \cite{pope,drazin_2003}).
The square cylinder is frequently used as a canonical test for numerical schemes and has been extensively studied in the past \cite{okajima82,sohankar98,darekar01,shahbazi07}, while the flow over a NACA 65(1)-412 cambered airfoil at incidence is a relevant case encountered in engineering applications, e.g. in turbomachinery \cite{nelson16}. 

On the basis of the examples introduced above, we show that the split form DGSEM produces numerically stable results of Navier-Stokes flows where the standard scheme fails. 
While the computations converge to the same solutions for high enough polynomial orders, the robustness of the kinetic energy conserving split form for simulations of marginally resolved flows makes it the primary choice for implicit LES computations.

%

The governing equations and the numerical model are outlined in Section \ref{methodology}. In Section \ref{results}, we present the setup and results of our computations. A summary and conclusion are presented at the end of this paper in Section \ref{conclusion}. 

\section{Methodology} \label{methodology}
\subsection*{Conservation Laws}
We compute solutions to the compressible Navier-Stokes equations, which can be written in non-dimensional form as the system of equations 
\begin{equation}\label{eq:n-s}
\partial_t\mathbf{U} + \nabla\cdot\mathbf{F} = 0.
\end{equation}
In \eqref{eq:n-s}, $\mathbf{U}$ represents the vector of the conserved variables,
\begin{equation}
\mathbf{U} = \left[\,\rho \quad \rho u \quad \rho v \quad \rho w \quad \rho e\,\right]^T.
\end{equation}
The flux vector $\mathbf{F}$ is split into an advective (superscript \textit{a}) and a viscous part (superscript \textit{v}),
\begin{equation} \label{eq:flux}
\nabla\cdot\mathbf{F} = \partial_x\mathbf{F}^a + \partial_y\mathbf{G}^a + \partial_z\mathbf{H}^a - \frac{1}{Re_f}\left(\partial_x\mathbf{F}^v + \partial_y\mathbf{G}^v + \partial_z\mathbf{H}^v\right),
\end{equation}
where
\begin{align}
\begin{split}
\mathbf{F}^a = \left[\,
\rho u \quad p {+} \rho u^2 \quad \rho u v \quad \rho u w \quad u(\rho e {+} p)
\,\right]^T,
\\
\mathbf{G}^a = \left[\,
\rho v \quad \rho v u \quad p {+} \rho v^2 \quad \rho v w \quad v(\rho e {+} p)
\,\right]^T,
\\
\mathbf{H}^a = \left[\,
\rho w \quad \rho w u \quad \rho w v \quad p {+} \rho w^2 \quad w(\rho e {+} p)
\,\right]^T,
\end{split}
\end{align}
\begin{align}
\begin{split}
\mathbf{F}^v = \left[\,
0 \quad \tau_{xx} \quad \tau_{yx} \quad \tau_{zx} \quad u \tau_{xx} {+} v \tau_{yx} {+} w \tau_{zx} {+} \frac{\kappa}{\left(\gamma -1\right)Pr M_f^2} T_x
\,\right]^T,
\\
\mathbf{G}^v = \left[\,
0 \quad \tau_{xy} \quad \tau_{yy} \quad \tau_{zy} \quad u \tau_{xy} {+} v \tau_{yy} {+} w \tau_{zy} {+} \frac{\kappa}{\left(\gamma -1\right)Pr M_f^2} T_y
\,\right]^T,
\\
\mathbf{H}^v = \left[\,
0 \quad \tau_{xz} \quad \tau_{yz} \quad \tau_{zz} \quad u \tau_{xz} {+} v \tau_{yz} {+} w \tau_{zz} {+} \frac{\kappa}{\left(\gamma -1\right)Pr M_f^2} T_z
\,\right]^T.
\end{split}
\end{align}
$\rho$, $u$, $v$, $w$, $p$, and $T$ are the density, velocities, pressure, and temperature respectively. The specific total energy is $\rho e = p/(\gamma-1)+\frac{1}{2}\rho(u^2+v^2+w^2)$ and the system is closed by the equation of state,
\begin{equation}
p = \frac{\rho T}{\gamma M_f^2}.
\end{equation}

All quantities are non-dimensionalized with respect to a problem specific reference length, velocity, density, and temperature yielding the non-dimensional Reynolds number, $Re_f$ and Mach number, $M_f$.

\subsection*{The Discontinuous Galerkin Spectral Element Method}
We approximate the system, \eqref{eq:n-s}, with a 
discontinuous Galerkin spectral element method (DGSEM). Details can be found in \cite{kopriva,Gassner2018} and we only provide a short summary here. 

The physical domain is subdivided into hexahedral elements, each of which is mapped from the reference element, $E = [-1,1]^{3}$ by a transformation $\vec x = \vec X\left(\xi,\eta,\zeta\right)$. Under the transformation, the reference space equations become 
\begin{equation}
\tilde {\mathbf U}_{t}+\nabla_{\xi}\cdot\tilde {\mathbf F}=0,
\label{eq:MappedEqns}
\end{equation}
where $\tilde {\mathbf U} = J\mathbf U$, $J$ is the transformation Jacobian, and $\tilde {\mathbf F}$ is the contravariant flux.
  
The DGSEM approximates the conserved variables and the contravariant fluxes as polynomials of arbitrary order \textit{N} within each element. We approximate the vector $\tilde{\mathbf{U}}$ as
\begin{equation}
\tilde{\mathbf{U}}_N = \sum_{i=0}^{N}\sum_{j=0}^{N}\sum_{k=0}^{N}\left(\tilde{\mathbf{U}}_{N}\right)_{i,j,k}\ell_i(\xi)\ell_j(\eta)\ell_k(\zeta),
\end{equation}
where the Lagrange interpolating polynomials, $\ell_{i}(\xi)$ are 
\begin{equation}
\ell_i(\xi) = \prod_{\substack{n=0 \\ n\neq i}}^N \frac{\xi-\xi_n}{\xi_i-\xi_n},
\end{equation}
and similarly for $\ell_{j}\left(\eta\right)$ and $\ell_{k}\left(\zeta\right)$.
The nodes $\xi_i$, $\eta_{j}$, and $\zeta_{k}$ are chosen to be the nodes of a Gauss quadrature. 

\subsubsection*{Weak Formulation}
The approximation satisfies a weak form the conservation law, constructed by taking the inner product of \eqref{eq:MappedEqns} with a test function $\phi$,
\begin{equation}
\int_{E}\left(\partial_t  \tilde{\mathbf{U}} + \nabla\cdot \tilde{\mathbf{F}}\right)\phi\; \mathrm{d}\mathbf{\xi} = 0,
\end{equation}
and integrating by parts
\begin{equation} \label{eq:ipb}
\int_{E}\partial_t \tilde{\mathbf{U}}\phi\; \mathrm{d}\mathbf{\xi} + \int_{\partial E}\tilde{\mathbf{F}}\cdot\mathbf{n}\phi\; \mathrm{d}S - \int_{E} \tilde{\mathbf{F}}\cdot\nabla\phi \; \mathrm{d}\mathbf{\xi} = 0.
\end{equation}

DG approximations do not require the solution to be continuous at the interface, and elements are coupled through the boundary flux in \eqref{eq:ipb}. We replace $\tilde{\mathbf{F}}$ with a numerical flux $\tilde{\mathbf{F}}^*\left(\tilde{\mathbf{U}}^{L},\tilde{\mathbf{U}}^{R}\right)$, which depends only on the solutions on the left and right of the interface between two elements, and is computed through a Riemann solver, e.g. the upwinding scheme by Roe \cite{roe}.

The integrals in \eqref{eq:ipb} are approximated with a Gauss quadrature of $N+1$ nodes, and two choices have been commonly used. The first is the Legendre-Gauss (LG) quadrature, which approximates the integral exactly for polynomial integrands of order $2N+1$ or less, but whose nodes do not include endpoints. The second is the Legendre-Gauss-Lobatto (LGL) quadrature, whose nodes include endpoints, but is only exact for polynomial integrands of order $2N-1$ or less. 
For a more detailed discussion of the differences between LG and LGL quadrature we refer to Gassner and Kopriva \cite{gassner2011}.

By replacing the integrals in \eqref{eq:ipb} with quadrature, and choosing $\phi = \ell_{i}(\xi)\ell_{j}(\eta)\ell_{k}(\zeta)$, the flux derivatives for the Gauss-Lobatto version become 
\begin{align}
\label{eq:weak_fluxes}
\begin{split}
\left.\partial_\xi\tilde{\mathbf{F}}\right|_{ijk} \approx \left(\delta_{iN}\tilde{\mathbf{F}}^{*}_{Njk} - \delta_{i0}\tilde{\mathbf{F}}^{*}_{0jk}\right) + \sum_{m=0}^N \hat{D}_{im}\tilde{\mathbf{F}}_{mjk},
\\
\left.\partial_\eta\tilde{\mathbf{G}}\right|_{ijk} \approx \left(\delta_{jN}\tilde{\mathbf{G}}^{*}_{iNk} - \delta_{j0}\tilde{\mathbf{G}}^{*}_{i0k}\right) + \sum_{m=0}^N \hat{D}_{jm}\tilde{\mathbf{G}}_{imk},
\\
\left.\partial_\zeta\tilde{\mathbf{H}}\right|_{ijk} \approx \left(\delta_{kN}\tilde{\mathbf{H}}^{*}_{ijN} - \delta_{k0}\tilde{\mathbf{H}}^{*}_{ij0}\right) + \sum_{m=0}^N \hat{D}_{km}\tilde{\mathbf{H}}_{ijm},
\end{split}
\end{align}
where $\hat{D}_{ij} = -D_{ji}w_j/w_i$ and $D_{ij} = \ell_j'(\xi_i), i, j = 0,\dots, N$ is the derivative matrix. 

\subsubsection*{Formulations in strong and split form}
Integrating \eqref{eq:ipb} by parts one more time lets us rewrite \eqref{eq:ipb} in what is known as the strong form
\begin{equation} \label{eq:ipb2}
\int_{E}\partial_t \tilde{\mathbf{U}}\phi\; \mathrm{d}\mathbf{\xi} + \int_{\partial E}\left(\tilde{\mathbf{F}}^*-\tilde{\mathbf{F}}\right)\cdot\mathbf{n}\phi\; \mathrm{d}S + \int_{E}\nabla\cdot\tilde{\mathbf{F}}\phi \; \mathrm{d}\mathbf{\xi} = 0.
\end{equation}

Again, we replace the integrals with quadrature and choose $\phi = \ell_{i}(\xi)\ell_{j}(\eta)\ell_{k}(\zeta)$, so that for the Gauss-Lobatto version the flux derivatives in \eqref{eq:ipb2} become \cite{kopriva}
\begin{align}
\label{eq:strong_fluxes}
\begin{split}
\left.\partial_\xi\tilde{\mathbf{F}}\right|_{ijk} \approx \left(\delta_{iN}\left[\tilde{\mathbf{F}}^{*}-\tilde{\mathbf{F}}\right]_{Njk} - \delta_{i0}\left[\tilde{\mathbf{F}}^{*}-\tilde{\mathbf{F}}\right]_{0jk}\right) + \sum_{m=0}^N D_{im}\tilde{\mathbf{F}}_{mjk},
\\
\left.\partial_\eta\tilde{\mathbf{G}}\right|_{ijk} \approx \left(\delta_{jN}\left[\tilde{\mathbf{G}}^{*}-\tilde{\mathbf{G}}\right]_{iNk} - \delta_{j0}\left[\tilde{\mathbf{G}}^{*}-\tilde{\mathbf{G}}\right]_{i0k}\right) + \sum_{m=0}^N D_{jm}\tilde{\mathbf{G}}_{imk},
\\
\left.\partial_\zeta\tilde{\mathbf{H}}\right|_{ijk} \approx \left(\delta_{kN}\left[\tilde{\mathbf{H}}^{*}-\tilde{\mathbf{H}}\right]_{ijN} - \delta_{k0}\left[\tilde{\mathbf{H}}^{*}-\tilde{\mathbf{H}}\right]_{ij0}\right) + \sum_{m=0}^N D_{km}\tilde{\mathbf{H}}_{ijm},
\end{split}
\end{align}
where $D_{ij} = \ell_j'(\xi_i), i, j = 0,\dots, N$ is the derivative matrix. Note that the forms \eqref{eq:weak_fluxes} and \eqref{eq:strong_fluxes} are algebraically equivalent \cite{kopriva2010}.

The non-linearity of the inviscid Euler fluxes introduces aliasing errors when the fluxes are approximated by polynomials, which can lead to instability. Gassner and collaborators \cite{gassner16,Gassner2018} showed that through the SPB property of the derivative operator when using the LGL points, the volume term contributions in \eqref{eq:strong_fluxes} can be rewritten as
\begin{align}
\label{eq:split_form}
\begin{split}
\left.\partial_\xi\tilde{\mathbf{F}}\right|_{ijk} \approx \left(\delta_{iN}\left[\tilde{\mathbf{F}}^{*}-\tilde{\mathbf{F}}\right]_{Njk} - \delta_{i0}\left[\tilde{\mathbf{F}}^{*}-\tilde{\mathbf{F}}\right]_{0jk}\right) + 2\sum_{m=0}^N D_{im}\tilde{\mathbf{F}}^{\#}_{(i,m),j,k},
\\
\left.\partial_\eta\tilde{\mathbf{G}}\right|_{ijk} \approx \left(\delta_{jN}\left[\tilde{\mathbf{G}}^{*}-\tilde{\mathbf{G}}\right]_{iNk} - \delta_{j0}\left[\tilde{\mathbf{G}}^{*}-\tilde{\mathbf{G}}\right]_{i0k}\right) + 2\sum_{m=0}^N D_{jm}\tilde{\mathbf{G}}^{\#}_{i,(j,m),k},
\\
\left.\partial_\eta\tilde{\mathbf{H}}\right|_{ijk} \approx \left(\delta_{kN}\left[\tilde{\mathbf{H}}^{*}-\tilde{\mathbf{H}}\right]_{ijN} - \delta_{k0}\left[\tilde{\mathbf{H}}^{*}-\tilde{\mathbf{H}}\right]_{ij0}\right) + 2\sum_{m=0}^N D_{km}\tilde{\mathbf{H}}^{\#}_{i,j,(k,m)}.
\end{split}
\end{align}
so that the scheme becomes entropy or energy conserving, depending on the choice of the new two-point fluxes $\mathbf{F}^{a,\#}$, $\mathbf{G}^{a,\#}$, and $\mathbf{H}^{a,\#}$, and the numerical surface fluxes. In this work, we choose the kinetic energy conserving split form by Pirozzoli \cite{pirozzoli}
\begin{equation} \label{eq:PI}
\mathbf{F}^{a,\#} = \begin{bmatrix}
\{\{\rho\}\} \{\{u\}\} \\ \{\{\rho\}\} \{\{u\}\}^2 {+} \{\{p\}\} \\ \{\{\rho\}\} \{\{u\}\} \{\{v\}\} \\ \{\{\rho\}\} \{\{u\}\} \{\{w\}\} \\ \{\{\rho\}\} \{\{u\}\} \{\{h\}\}
\end{bmatrix},
\enspace
\mathbf{G}^{a,\#} = \begin{bmatrix}
\{\{\rho\}\} \{\{v\}\} \\ \{\{\rho\}\} \{\{u\}\} \{\{v\}\} \\ \{\{\rho\}\} \{\{v\}\}^2 {+} \{\{p\}\} \\ \{\{\rho\}\} \{\{v\}\} \{\{w\}\} \\ \{\{\rho\}\} \{\{v\}\} \{\{h\}\}
\end{bmatrix},
\enspace.
\mathbf{H}^{a,\#} = \begin{bmatrix}
\{\{\rho\}\} \{\{w\}\} \\ \{\{\rho\}\} \{\{u\}\} \{\{w\}\} \\ \{\{\rho\}\} \{\{v\}\} \{\{w\}\} \\ \{\{\rho\}\} \{\{w\}\}^2 {+} \{\{p\}\} \\ \{\{\rho\}\} \{\{w\}\} \{\{h\}\}
\end{bmatrix},
\end{equation}
with the notation $\{\{a\}\}_{im} := \frac{1}{2}(a_i + a_m)$. 

The viscous stresses are computed in the standard formulation $\mathbf{F}^{v,\#} = \{\{\mathbf{F}^v\}\}$, $\mathbf{G}^{v,\#} = \{\{\mathbf{G}^v\}\}$, and $\mathbf{H}^{v,\#} = \{\{\mathbf{H}^v\}\}$ and
the total fluxes are $\mathbf{F}^{\#} = \mathbf{F}^{a,\#}-\mathbf{F}^{v,\#}$, with the contravariant forms follow from $\tilde{\mathbf{F}}^{\#} =\left[\mathbf{F}^{\#},\mathbf{G}^{\#},\mathbf{H}^{\#} \right]\cdot\nabla\xi$, see \cite{kopriva,Gassner2018}. Formulations for $\tilde{\mathbf{G}}^{\#}$ and $\tilde{\mathbf{H}}^{\#}$ are obtained similarly.

It was also shown in \cite{gassner16} that the scheme is kinetic energy conserving only if a central Riemann solver, e.g. the Lax-Friedrichs numerical flux without additional stabilization terms, is chosen as the numerical flux. Additional dissipation can be added through the appropriate choice of Riemann solver to increase robustness or for implicit large eddy simulations \cite{flad17}.

To advance the solution in time, the system of equations \eqref{eq:n-s} is integrated with an explicit low-dispersion 5-stage 4th-order Runge-Kutta scheme. 

\section{Results and Discussion} \label{results}
In the this section, we present the setup and computed results for the inviscid Taylor-Green vortex, the square cylinder flow, the plane jet and the flow over a NACA 65(1)-412 airfoil.
Each problem is simulated using the standard DGSEM, which supports both Gauss and Gauss-Lobatto nodes, and the kinetic energy conserving formulation of the split form (SF) with Gauss-Lobatto points.
A Courant-Fridrichs-Lewy (CFL) number of 0.8 is chosen unless otherwise stated.

\subsection*{Taylor-Green Vortex}
\begin{figure}[htp]
	\centering
	\begin{subfigure}[b]{0.5\textwidth}
 	\centering
	\includegraphics[height=13em]{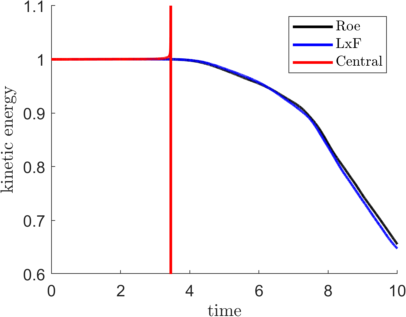}
	\caption{Gauss nodes \& standard DG}
	\end{subfigure}\hfill
	\begin{subfigure}[b]{0.5\textwidth}
 	\centering
	\includegraphics[height=13em]{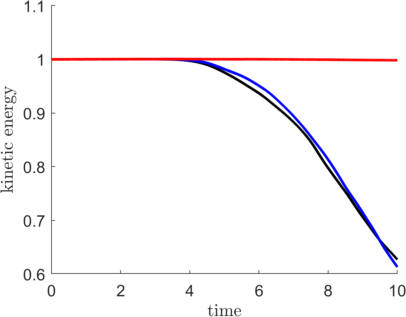}
	\caption{Gauss-Lobatto nodes \& split form DG}
	\end{subfigure}
	\caption{Integrated kinetic energy over time of the Taylor-Green vortex in 3D with different Riemann solvers (Roe, Lax-Friedrichs (LxF), Central without dissipation). (a) Standard form with LG nodes. (b) Split form with LGL nodes.}
	\label{fig:taylorgreen}
\end{figure}

The implementation of the kinetic-energy preserving inviscid fluxes into the DGSEM code is verified with the inviscid Taylor-Green vortex in three dimensions, as it has been done by Gassner et al. \cite{gassner16}. 16\textsuperscript{3} elements are used in combination with a uniform polynomial order of \textit{N} = 3. The kinetic energy $k=(u^2+v^2+w^2)/2$ is integrated over the domain at each time step and plotted in Figure \ref{fig:taylorgreen}. Two cases are considered: the standard form using LG nodes and the split form with LGL nodes, where the interface fluxes are computed with an upwinding Roe solver (Roe), a central Lax-Friedrichs solver with a stabilization term (LxF), and a central Riemann solver without dissipation (Central). 

The kinetic energy as a function of time is shown in Figure \ref{fig:taylorgreen}. As expected, the standard form without dissipation at the interfaces is unstable. Upwinding or dissipative schemes are stable for this problem, but decrease the total kinetic energy with time (Fig. \ref{fig:taylorgreen}a). The split form approximation conserves the kinetic energy if a central Riemann solver without stabilization terms is used, while the Roe and Lax-Fiedrichs solvers add numerical viscosity and decrease the total kinetic energy (Fig. \ref{fig:taylorgreen}b), just as in the standard form. It should be noted that the split form approximation is not unconditionally stable, as the same test case crashes at higher polynomial orders \cite{gassner16}.
In the following test problems, the upwinding Riemann solver of Roe is used.

\subsection*{Plane Jet}
Fluid jets are naturally unstable shear flows that show exponential growth to perturbations and eventually transition to turbulence \cite{drazin_2003}.
A high-order scheme can accurately capture these instabilities, but its numerical robustness might be challenged by the developing turbulence in this flow. 
We simulate the two-dimensional flow of a strong jet at a Reynolds number of $Re_h$ = 27,300, based on the jet width $h$ and the difference between centerline and coflow velocities.
The computational domain consists of 3,200 elements over -10$h$ $\leq$ y $\leq$ 10$h$ and 0 $\leq$ x $\leq$ 40$h$.

A top-hat inflow velocity profile is set at the left boundary, with the commonly used hyperbolic tangent function
\begin{equation}
u = \frac{U_1+U_2}{2}+\frac{U_1-U_2}{2}\tanh\left(\frac{y}{2\Theta}\right).
\end{equation}
Here, $U_1$ is the centerline velocity and $U_2$ is the coflow velocity. 
We set $U_1$ = 1 and $U_2$ = 0.09, which results in a velocity ratio $\eta$ = $\Delta U/(U_1+U_2)$ = 0.83 between the flows.
The parameter $\Theta$ is the inflow momentum thickness and is set to $\Theta=h/20$, in accordance with the setup used by Stanley et al. \cite{stanley02}. At the other boundaries, a free-stream pressure condition is prescribed. A buffer layer of thickness 3$h$ gradually damps the solution to the uniform free-stream value to avoid spurious reflections from the boundaries, as described by Stanescu et al. \cite{stanescu99}. Similar approaches have been used by Jacobs et al. \cite{JKM03} and Rasetarinera et al. \cite{Rasetarineraetal2001}.
We compute the flow using polynomial orders of \textit{N} = [1, 3, 5, 7, 9] and extract the temporal spectrum of the turbulent kinetic energy at the center line at $x/h$ = 20. The velocity history and average flow field is recorded over 1,000 convective time units.

\begin{figure}[htp]
	\centering
	\includegraphics[height=12em]{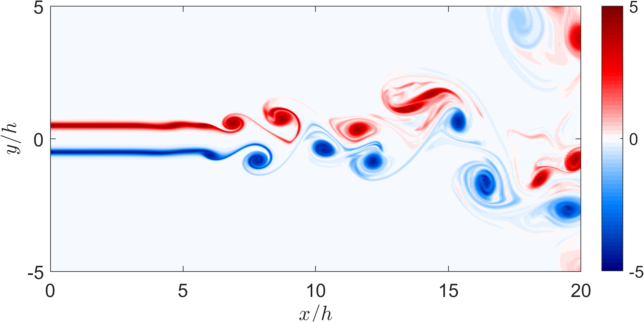}
	\caption{Instantaneous vorticity of the jet flow (\textit{N} = 9).}
	\label{fig:jet_vort}
\end{figure}

Figure \ref{fig:jet_vort} shows instantaneous vorticity contours of the jet flow. Driven by the Kelvin-Helmholtz instability, vortices form in the shear layer and break up the jet.

Table \ref{tab:jet} summarizes the numerical robustness of the approximations for this flow. The standard form with Gauss-Lobatto nodes is numerically unstable at all polynomial orders, while the split form remains stable for all cases. The standard form with Gauss nodes only crashes for this flow if a polynomial order of \textit{N} = 1 is used. 
\begin{table}
\caption{Numerical Stability of the plane jet. \checkmark = stable, x = unstable.}
\centering
\begin{tabular}{l@{\hskip 2em} l@{\hskip 2em} | c@{\hskip 1em} c@{\hskip 2em} c@{\hskip 2em} c@{\hskip 2em} c}
\hline
Node type & Flux method & \textit{N} = 1 & \textit{N} = 3 & \textit{N} = 5 & \textit{N} = 7 & \textit{N} = 9 \\
\hline
LG  & Standard    & x          & \checkmark & \checkmark & \checkmark & \checkmark \\
LGL & Standard    & x          & x          & x          & x          & x          \\
LGL & Split form  & \checkmark & \checkmark & \checkmark & \checkmark & \checkmark \\
\end{tabular}
\label{tab:jet}
\end{table}

\begin{figure}[htp]
	\centering
	\begin{subfigure}[b]{0.5\textwidth}
 	\centering
	\includegraphics[height=14em]{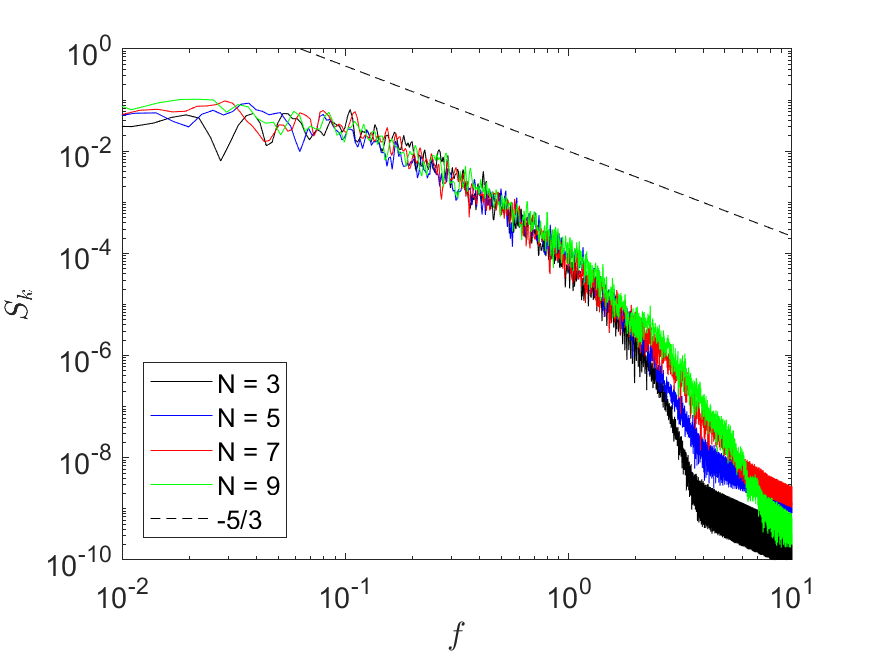}
	\caption{Gauss nodes \& standard DG}
	\end{subfigure}\hfill
	\begin{subfigure}[b]{0.5\textwidth}
 	\centering
	\includegraphics[height=14em]{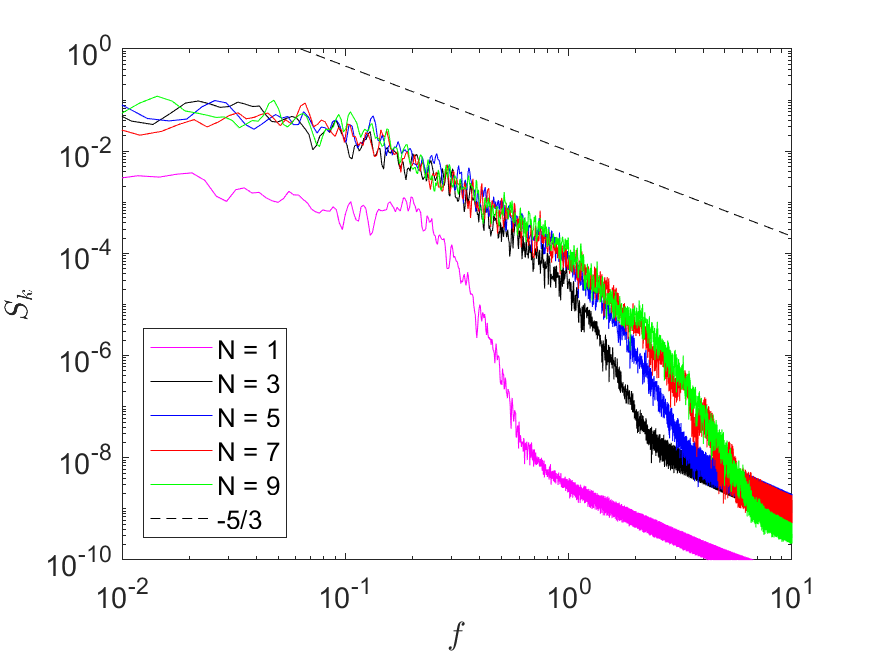}
	\caption{Gauss-Lobatto nodes \& split form DG}
	\end{subfigure}
	\caption{Temporal power spectra of the turbulent kinetic energy for (a) standard form with LG nodes and (b) split form with LGL nodes.}
	\label{fig:psd}
\end{figure}
Figure \ref{fig:psd} shows the power spectra of the turbulent kinetic energy $S_k$ computed from the velocity history at the center line at $x/h$ = 20. Although the split form gives a result for a polynomial order of \textit{N} = 1, the large deviation of the spectrum to the higher-order solutions renders the physical meaning of this solution irrelevant. All spectra obtained from higher-order simulations initially show a -5/3 decay, but an earlier start of the dissipation range at lower resolution indicates increased numerical dissipation at orders \textit{N} = 3 and \textit{N} = 5. Although the spectra for \textit{N} = 9 are nearly identical for the standard and the split form, the lower-order results indicate higher accuracy when using the Gauss nodes, as the solution obtained with Gauss-Lobatto nodes is overly dissipated at higher frequencies. This result is the effect of underintegration when using Gauss-Lobatto nodes, which acts as a modal filter on the highest modes, as explained by Gassner and Kopriva \cite{gassner2011}.

\begin{figure}[htp]
	\centering
	\begin{subfigure}[b]{0.5\textwidth}
 	\centering
	\includegraphics[height=12em]{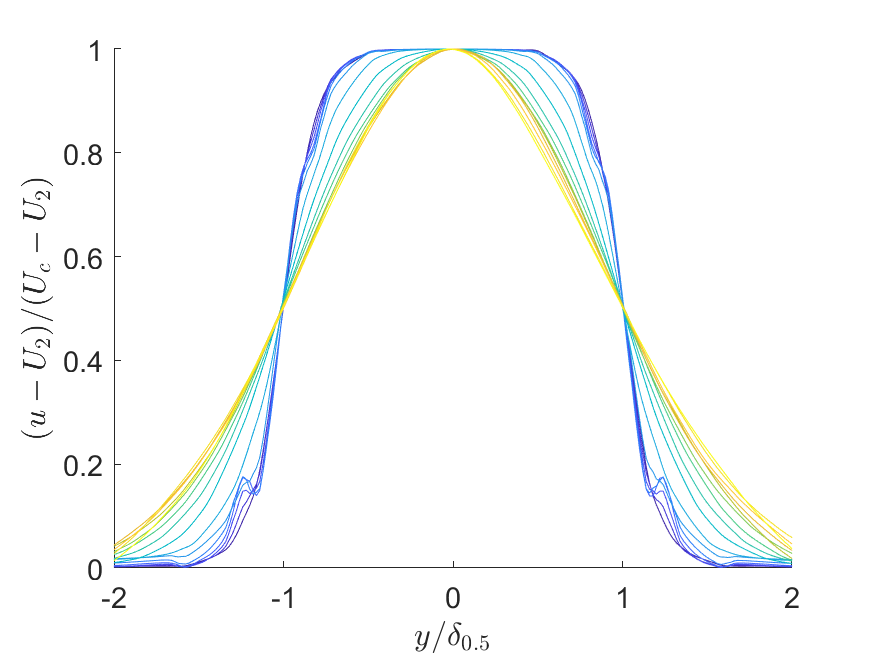}
	\caption{LG, \textit{N} = 3}
	\end{subfigure}\hfill
	\begin{subfigure}[b]{0.5\textwidth}
 	\centering
	\includegraphics[height=12em]{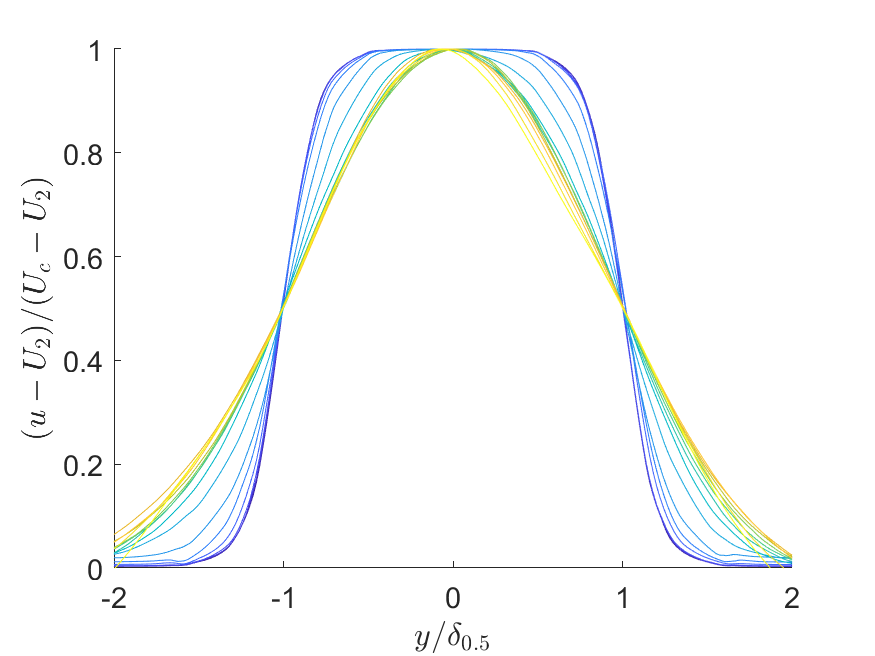}
	\caption{LGL-SF, \textit{N} = 3}
	\end{subfigure}
	\vskip\baselineskip
	\begin{subfigure}[b]{0.5\textwidth}
 	\centering
	\includegraphics[height=12em]{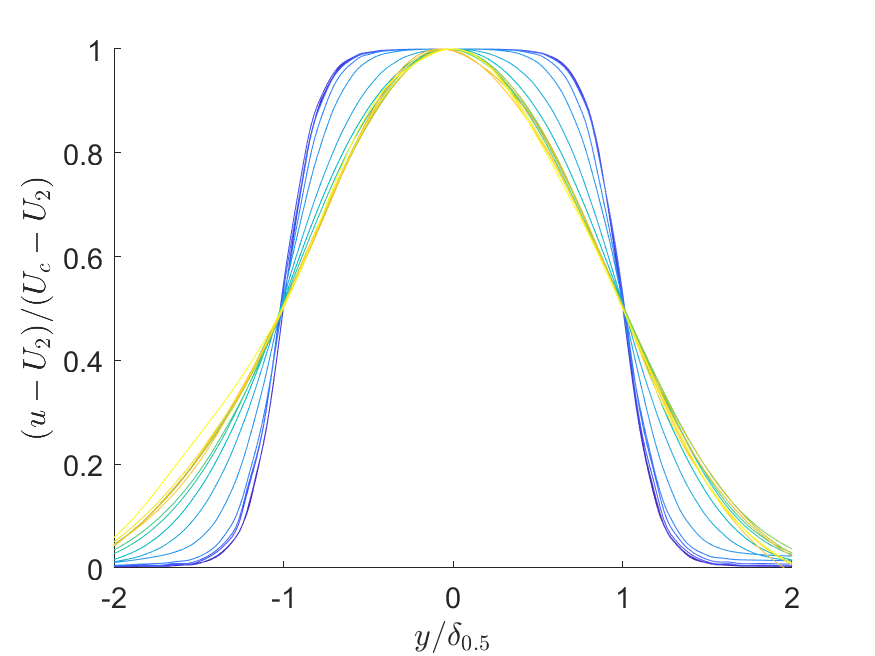}
	\caption{LG, \textit{N} = 5}
	\end{subfigure}\hfill
	\begin{subfigure}[b]{0.5\textwidth}
 	\centering
	\includegraphics[height=12em]{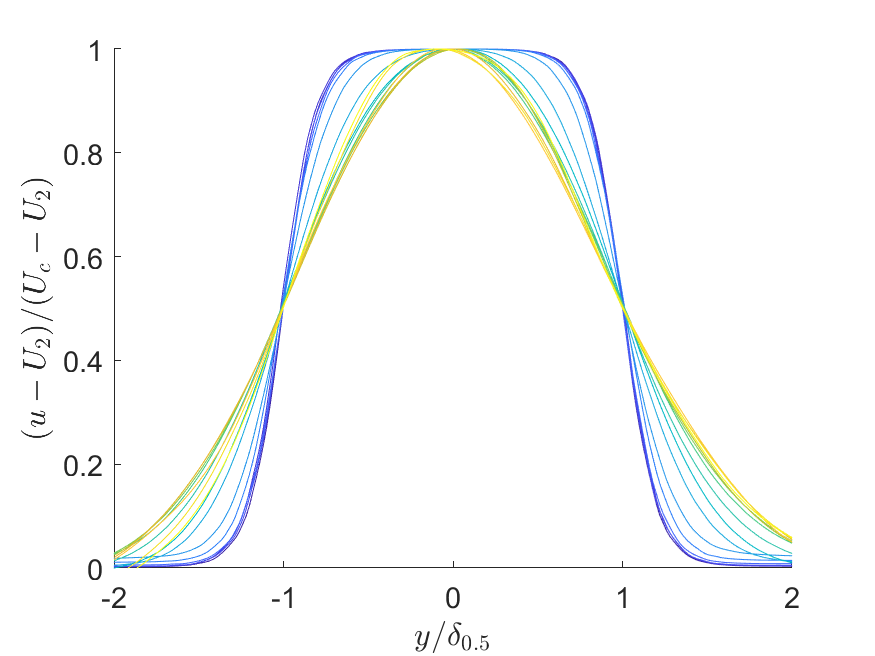}
	\caption{LGL-SF, \textit{N} = 5}
	\end{subfigure}
	\caption{Normalized velocity profiles at different streamwise locations. The color indicates the location from $x/h$ = 0 (blue) to $x/h$ = 15 (yellow). LG = Legendre-Gauss nodes (standard DG formulation). LGL-SF = Legendre-Gauss-Lobatto nodes \& split form DG).}
	\label{fig:uprof}
\end{figure}

Figure \ref{fig:uprof} shows the normalized velocity profile over the jet half-width for different streamwise locations for polynomial orders of \textit{N} = 3 and \textit{N} = 5. While the initial top-hat profile dominates close to the inlet, the well-known self-similar solution is obtained further away. 
The low-order approximation obtained with Gauss nodes shows a spurious oscillation in the velocity profile close to the inlet (Fig. \ref{fig:uprof}a), while the result of the LGL solution does not. At higher order (\textit{N} = 5), the profiles are similar and no spurious oscillations are present in the Gauss approximation. 

As noted by Gassner and Kopriva \cite{gassner2011}, using Gauss-Lobatto nodes reduces the stiffness of the problem and can allow larger time steps. For the jet flow, we summarize the maximum stable time step sizes in Table \ref{tab:dt} and show that simulations using the split form allow for a 66\% (\textit{N} = 5) to 32\% (\textit{N} = 9) increase in the step size $\Delta t$ over the standard DG formulation. The additional work required to compute the split form, however, results in an overhead of 64\% (\textit{N} = 5) to 150\% (\textit{N} = 9) per time step on the machine tested (Intel E5520), but has been observed to vary significantly between different computer systems.
\begin{table}
\caption{Maximum stable convective time step size $\Delta t \cdot 10^{-3}$, x = unstable.}
\centering
\begin{tabular}{l@{\hskip 2em} l@{\hskip 2em} | c@{\hskip 2em} c@{\hskip 2em} c}
\hline
Node type & Flux method & \textit{N} = 5 & \textit{N} = 7 & \textit{N} = 9 \\
\hline
LG  & Standard    & 2.8 & 1.9 & 1.2 \\
LGL & Standard    & x   & x   & x   \\
LGL & Split form  & 4.7 & 2.6 & 1.6 \\
\end{tabular}
\label{tab:dt}
\end{table}

Although the spectra indicate superior accuracy of the Gauss nodes for under-resolved computations, the results must be examined carefully, as spurious oscillations might be present. In such a case, the underintegrated solution of the Lobatto-Gauss nodes and its higher numerical diffusion might be beneficial, as such approximations should rather be considered implicit LES than DNS. 
This test case also highlights the effect of the kinetic energy conserving volume fluxes on the robustness of the scheme, as the standard form with LGL nodes crashes for all tested polynomial orders while the split form is numerically stable. 

\subsection*{Square Cylinder Flow}
The canonical flow over a square cylinder at a Reynolds number of \textit{Re} = 100 is chosen because of the numerous data sets available in the literature \cite{JKM03,okajima82,sohankar98,darekar01,shahbazi07}, which makes it a good candidate for the verification of the numerical method. The Strouhal numbers reported in the literature are summarized in Table \ref{tab:strouhal_lit}.
\begin{table}
\caption{Strouhal numbers of square cylinder flow at \textit{Re} = 100 (literature).}
\centering
\begin{tabular}{l@{\hskip 6em} r}
\hline
Source                                & \textit{St} \\
\hline
Okajima \cite{okajima82}			  & 0.141-0.145 \\
Sohankar et al. \cite{sohankar98}  	  & 0.144-0.146 \\
Darekar and Sherwin \cite{darekar01}  & 0.145       \\
Shahbazi et al. \cite{shahbazi07}     & 0.145       \\
\end{tabular}
\label{tab:strouhal_lit}
\end{table}

The computational domain consists of 3,350 quadrilateral elements and has a blockage of 2.3\%, matching previous computations in literature \cite{darekar01,shahbazi07}. Three polynomial orders of \textit{N} = [1, 2, 4] are used in the simulations. The cylinder walls are approximated with no-slip adiabatic walls and free stream conditions are applied at the outer boundaries. A Mach number of \textit{M} = 0.1 renders compressibility effects negligible. 
For all cases, the Strouhal number $St$ = $fL/U$, based on the frequency of the lift coefficient, is computed over 400 convective time units after quasi-steady state is reached.

\begin{figure}[htp]
	\centering
	\includegraphics[height=11em]{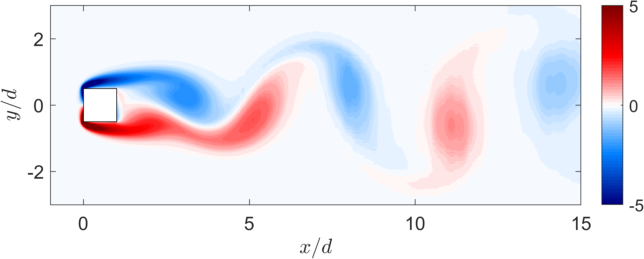}
	\caption{Instantaneous vorticity plot (\textit{N} = 4).}
	\label{fig:cylinder_vort}
\end{figure}

A plot of the instantaneous vorticity in Figure \ref{fig:cylinder_vort} shows the shedding of counter-rotating vortices from the square cylinder into a Von-Karman vortex street. This flow pattern is well known and has been observed in experiments, e.g. by Okajima \cite{okajima82}.

The Strouhal numbers computed from spectral analysis of the lift coefficient are summarized in Table \ref{tab:strouhal} and show convergence to \textit{St} = 0.145 at a polynomial of \textit{N} = 2, independent of the quadrature nodes or form of the advective fluxes. This value is in excellent agreement with the Strouhal numbers reported in literature (Tab. \ref{tab:strouhal_lit}). 

All computations are numerically stable, but at a polynomial order of \textit{N} = 1, the flows are under-resolved and differ from the converged solution by 4\% for the Gauss and 7\% - 8\% for Gauss-Lobatto quadrature nodes. Given that the quadrature with Gauss-Lobatto nodes is only exact for polynomials of order $2N-1$, Gassner and Kopriva \cite{gassner2011} showed that underintegration with Gauss-Lobatto points can be interpreted as a modal filter and leads to damping of the highest modes. The higher accuracy of the Gauss quadrature translates to a smaller error of the Strouhal number for the under-resolved flow. Differences between the standard and the split form are minor, as the error for this problem appears to be dominated by the choice of quadrature nodes.

\begin{table}[htp]
\caption{Strouhal numbers of square cylinder flow at \textit{Re} = 100 (present).}
\centering
\begin{tabular}{l@{\hskip 2em} l@{\hskip 2em} c@{\hskip 2em} c}
\hline
Node type & Flux method & \textit{N} & \textit{St} \\
\hline
LG   & Standard    & 1 & 0.139 \\
LGL  & Standard    & 1 & 0.135 \\
LGL  & Split form  & 1 & 0.134 \\
LG   & Standard    & 2 & 0.145 \\
LGL  & Standard    & 2 & 0.145 \\
LGL  & Split form  & 2 & 0.145 \\
LG   & Standard    & 4 & 0.145 \\
LGL  & Standard    & 4 & 0.145 \\
LGL  & Split form  & 4 & 0.145 \\
\end{tabular}
\label{tab:strouhal}
\end{table}

\subsection*{Two-Dimensional Airfoil Flow}
The flow over a NACA 65(1)-412 airfoil is simulated at a Reynolds number based on the chord length of $Re_c$ = 20,000 and a Mach number of \textit{M} = 0.3. The airfoil is at 4$^{\circ}$ incidence.

The computational domain consists of 2,256 quadrilateral elements, where the boundaries are curved and fitted to a spline representing the airfoil's surface as described by Nelson et al. \cite{nelson16}. The outer boundaries of the domain are defined as free-stream boundaries while the airfoil surface is treated as a non-slip, adiabatic wall. Polynomial orders of \textit{N} = [3, 6, 12] are used in all elements, where \textit{N} = 12 has been found to give a grid converged solution. A more detailed description of the base flow is given by Nelson \cite{nelson16} and Klose et al. \cite{klose18}.
All computations are run until quasi-steady state is reached, and statistics are computed over 10 convective time units. 

\begin{figure}[htp]
	\centering
	\includegraphics[height=8em]{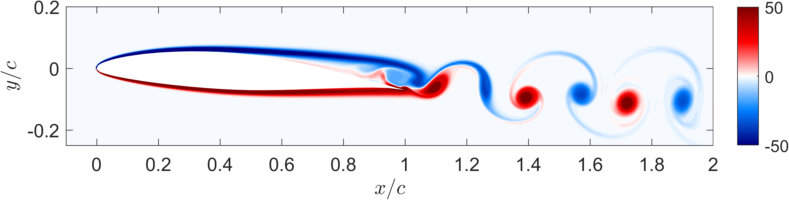}
	\caption{Instantaneous vorticity contours of flow over NACA 65(1)-412.}
	\label{fig:airfoil_vort}
\end{figure}

The vorticity plot in Figure \ref{fig:airfoil_vort} shows that the boundary layer separates mid-cord, enclosing a recirculation region on the airfoil's upper side. Periodic shedding of vortices lead to the formation of a Von-Karmann type vortex street in the wake.

The averaged lift and drag coefficients, as well as the Strouhal numbers are summarized in Table \ref{tab:airfoil_res}.
For a polynomial order of \textit{N} = 12, the averaged lift and drag coefficients are $\bar{C}_l$ = 0.442 for both LG and LGL-SF schemes and $\bar{C}_d$ = 0.0552 for the standard form with Gauss points and $\bar{C}_d$ = 0.0554 for the split form, resulting in a relative difference of <0.4\%. The Strouhal number based on the frequency of the lift coefficient is consistently at \textit{St} = 2.78 for all cases. The fact that no significant difference between the LG and LGL results are found is in accordance with the findings by Nelson et al., who reported convergence for a polynomial order of \textit{N} = 12 using LG points \cite{nelson16}. 

To evaluate the performance on under-resolved computations, we decrease the polynomial order to \textit{N} = 6 and \textit{N} = 3 and compare the results to the reference solution at \textit{N} = 12. 
At a polynomial order of \textit{N} = 3, the methods using Gauss-Lobatto quadrature nodes perform rather poorly, as the standard form is numerically unstable and the split form largely overestimates the drag (33\%, Tab. \ref{tab:airfoil_res}). Results obtained from the standard LG form are closer to the converged solution (<4\%), with the exception of the Strouhal number (21\% over-estimated).
\begin{table}
\caption{Aerodynamic data of airfoil flows at different polynomial orders.}
\centering
\begin{tabular}{l@{\hskip 2em} l@{\hskip 2em} l@{\hskip 2em} l@{\hskip 2em} l@{\hskip 2em} r}
\hline
Node type & Flux method & \textit{N} & $\bar{C}_l$ & $\bar{C}_d$ & \textit{St} \\
\hline
LG   & Standard    & 3  & 0.457 & 0.0530 & 3.37 \\
LGL  & Standard    & 3  & -     & -      & -    \\
LGL  & Split form  & 3  & 0.405 & 0.0732 & 2.30 \\
LG   & Standard    & 6  & 0.445 & 0.0561 & 2.69 \\
LGL  & Standard    & 6  & 0.415 & 0.0560 & 2.59 \\
LGL  & Split form  & 6  & 0.440 & 0.0565 & 2.75 \\
LG   & Standard    & 12 & 0.442 & 0.0552 & 2.78 \\
LGL  & Standard    & 12 & 0.442 & 0.0554 & 2.78 \\
LGL  & Split form  & 12 & 0.442 & 0.0554 & 2.78 \\
\end{tabular}
\label{tab:airfoil_res}
\end{table}

\begin{figure}[htp]
	\centering
	\begin{subfigure}[b]{0.5\textwidth}
 	\centering
	\includegraphics[height=14em]{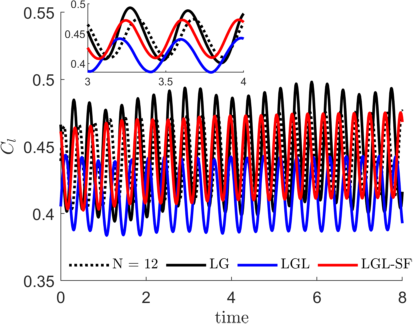}
	\caption{Lift coefficient}
	\end{subfigure}\hfill
	\begin{subfigure}[b]{0.5\textwidth}
 	\centering
	\includegraphics[height=14em]{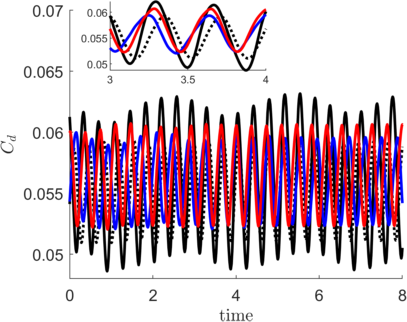}
	\caption{Drag coefficient}
	\end{subfigure}
	\caption{Lift (a) and drag (b) coefficients of underresolved flows for a polynomial orders of \textit{N} = 6. LG = Legendre-Gauss nodes (standard DG formulation), LGL = Legendre-Gauss-Lobatto nodes (standard DG formulation), LGL-SF = Legendre-Gauss-Lobatto nodes \& split form DG).}
	\label{fig:lift}
\end{figure}

As expected, increasing the polynomial order to \textit{N} = 6 considerably improves the quality of the results. The differences in the lift coefficient and Strouhal number of the split form are now below 1\% and only the drag is slightly over-estimated by about 2\%. The standard LG form matches lift and drag closely, but has a frequency error of 3\% with a spurious low-frequency component in the lift and drag force (Fig. \ref{fig:lift}). The standard form using LGL nodes gives the poorest results with larger errors in the lift and Strouhal number. 

Overall, the results show that for the marginally resolved airfoil flow at \textit{N} = 6, the split form yields the best results with the lift coefficient and Strouhal number matching the converged solution closely and only slightly larger deviations in the drag as compared to the standard DG scheme.
For strongly under-resolved flows, the standard formulation with LG nodes is advantageous, as the errors in lift and drag coefficients are much smaller than for simulations using LGL nodes. 
Again, this result is expected, as the Gauss-Lobatto quadrature underintegrates and thus is less accurate. The favorable dispersion relation of the Gauss-Lobatto nodes, as shown by Gassner and Kopriva \cite{gassner2011}, also gives a reasonable explanation for the better representation of the Strouhal number for under-resolved cases when using the LGL split form.

\FloatBarrier

\subsection*{Three-Dimensional Airfoil Flow}
Finally, we solve the three-dimensional flow over the NACA 65(1)-412 airfoil under an angle of attack of $\alpha$ = 10$^{\circ}$ at Reynolds number $Re_c$ = 20,000. 
The flow is characterized by a laminar separation bubble at the leading edge and subsequent transition to turbulence \cite{klose18}. 
The airfoil is extruded by half a cord length (0.5$c$) in the spanwise direction and periodic boundary conditions are applied to approximate an infinite wing, resulting in a total of 33,660 hexahedral elements. The computations are initialized by mapping a two-dimensional flow field uniformly in the spanwise direction.

Two cases are considered: the standard DG scheme with Gauss nodes and the kinetic energy stable split form with Gauss-Lobatto nodes. 
Polynomial orders of \textit{N} = 10 and \textit{N} = 12 in the region close to the airfoil are chosen, but the order is gradually lowered in the far field to decrease the computational costs.
A CFL number of 0.5 ensures that numerical instabilities do not arise from the time stepping scheme. 

\begin{table}
\caption{Numerical Stability of the 3D airfoil simulations at 10$^{\circ}$ incidence. \checkmark = stable, x = unstable.}
\centering
\begin{tabular}{l@{\hskip 2em} l@{\hskip 2em} | c@{\hskip 2em} c}
\hline
Node type & Flux method & \textit{N} = 10 & \textit{N} = 12 \\
\hline
LG  & Standard    & x          & x 			\\
LGL & Split form  & \checkmark & \checkmark \\
\end{tabular}
\label{tab:airfoil3D}
\end{table}

\begin{figure}[htp]
	\centering
	\begin{subfigure}[b]{0.5\textwidth}
 	\centering
	\includegraphics[height=12em]{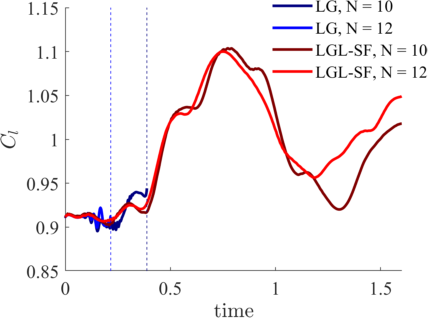}
	\caption{Lift coefficient}
	\end{subfigure}\hfill
	\begin{subfigure}[b]{0.5\textwidth}
 	\centering
	\includegraphics[height=12em]{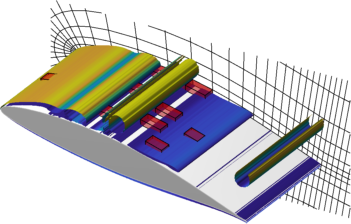}
	\caption{Iso-vorticity, \textit{t} = 0.2}
	\end{subfigure}
	\vskip\baselineskip
	\begin{subfigure}[b]{0.5\textwidth}
 	\centering
	\includegraphics[height=11em]{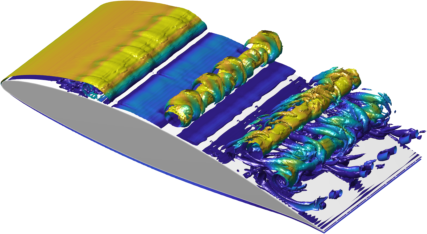}
	\caption{Iso-vorticity, \textit{t} = 1.6}
	\end{subfigure}\hfill
	\begin{subfigure}[b]{0.5\textwidth}
 	\centering
	\includegraphics[height=11em]{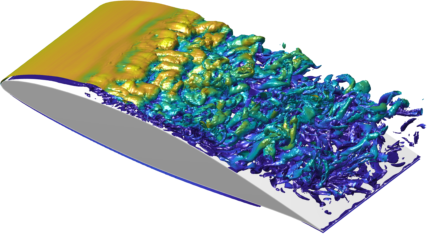}
	\caption{Iso-vorticity, \textit{t} = 18.1}
	\end{subfigure}
	\caption{(a) Lift coefficient for \textit{N} = 10 \& 12 and standard \& split form DG. Points of termination indicated by dashed lines. (b) -- (d): Iso-surfaces of the vorticity. Numerically unstable elements for the standard DG are highlighted in red.}
	\label{fig:airfoil3D}
\end{figure}

Only the split form DG scheme has the numerical robustness to compute this flow and produce results past the initial start-up phase, as summarized in Table \ref{tab:airfoil3D}. 
Figure \ref{fig:airfoil3D} (a) shows the lift coefficient over time for the different schemes and polynomial orders, where the dashed lines indicate the points of termination (crash).
Figures \ref{fig:airfoil3D} (b) -- (d) show iso-surfaces of the vorticity at $t$ = 0.2, $t$ = 1.6, and $t$ = 18.1 respectively and illustrate the flow transition to three dimensional turbulent structures. 
The numerically unstable elements in the standard DG scheme are highlighted in red in Figure \ref{fig:airfoil3D} (b).

Although the Reynolds number is moderate, the turbulent flow over the airfoil in Figure \ref{fig:airfoil3D} (d) illustrates the small-scale vortical structures and the need for a numerically robust scheme to capture them without filtering. Because the standard DG formulation crashes -- even with high-order approximations -- we see that transitional flows with high velocity gradients greatly benefit from the enhanced robustness of the split form DG scheme.

\section{Conclusion} \label{conclusion}
We have conducted a series of computations to evaluate the robustness and accuracy of the standard and a split form DG formulation for converged and marginally resolved viscous flows. The flows over a square cylinder, an airfoil and a plane jet are simulated for different polynomial orders, where each case is computed using the standard flux form with Legendre-Gauss and Legendre-Gauss-Lobatto quadrature nodes and the split form approximation of the advective fluxes with LGL nodes. 

It is shown that the Gauss DG scheme has a higher accuracy per point and increased numerical robustness over the the standard Gauss-Lobatto formulation and matches the converged solution more closely for the marginally resolved cylinder and jet flow cases. This result is in accordance with the work by Gassner \& Kopriva \cite{gassner2011}. 
There is, however, no proof of numerical stability, and errors in the non-linear terms result in the termination of the computation for under-resolved simulations if the dissipation (physical or interface) is not large enough. This is demonstrated by the three-dimensional Taylor-Green vortex and the airfoil simulation, where the Gauss DGSEM crashes, even for high polynomial orders.
Additionally, a spurious low frequency component in the lift and drag coefficient of the under-resolved two-dimensional airfoil simulation is present when using LG nodes.

The split form DGSEM, on the other hand, is provably stable in the sense that the kinetic energy is either conserved or dissipated, and generates solutions for marginally resolved flows that are inaccessible with the standard DG scheme. 
Although standard and split form DG approximations converge to the same solutions with increasing polynomial order, as shown for the two-dimensional cylinder, jet and airfoil flows, the split form DGSEM should be the primary choice for the computation of under-resolved flows, as it is robust and generates consistent results.

\section*{Acknowledgments}
We gratefully acknowledge funding by the Air Force Office of Scientific Research under FA9550-16-1-0392 of the Flow Control Program and from Solar Turbines. This work was supported by a grant from the Simons Foundation ($\#426393$, David Kopriva)


\bibliographystyle{unsrt}
\bibliography{bib}

\end{document}